# Quasiparticle band structure and optical properties of rutile GeO₂, an ultra-wide-band-gap semiconductor

Kelsey A. Mengle[1)], Sieun Chae[1)], and Emmanouil Kioupakis[1,a)]

[1]Department of Materials Science and Engineering, University of Michigan, Ann Arbor, 48109, United States

Rutile GeO₂ is a visible and near-ultraviolet-transparent oxide that has not been explored for semiconducting applications in electronic and optoelectronic devices. We investigate the electronic and optical properties of rutile GeO₂ with first-principles calculations based on density functional theory and many-body perturbation theory. Our band-structure calculations indicate a dipole-forbidden direct band gap at Γ with an energy of 4.44 eV and effective masses equal to $m_{e\perp}^* = 0.43\, m_0$, $m_{e\parallel}^* = 0.23\, m_0$, $m_{h\perp}^* = 1.28\, m_0$, and $m_{h\parallel}^* = 1.74\, m_0$. In contrast to the self-trapped hole polarons by lattice distortions in other wide-band-gap oxides that reduce the hole mobility, holes in rutile GeO₂ are delocalized due to their small effective mass. The first allowed optical transitions at Γ occur at 5.04 eV ($\vec{E} \perp \vec{c}$) and 6.65 eV ($\vec{E} \parallel \vec{c}$). We also evaluate the optical absorption coefficient and refractive index along both crystallographic directions. Our estimates for the exciton binding energies using the Bohr model are close to the reported experimental value. The ultra-wide-band-gap and light carrier effective masses of rutile GeO₂, coupled with its optical transparency in the visible and near UV are promising for applications in UV-transparent conductors and solar-blind photodetectors.

## I. INTRODUCTION

Germanium dioxide exists in multiple polytypes, such as the low-density α-quartz (trigonal) phase with tetrahedrally coordinated Ge atoms and the octahedrally coordinated rutile (tetragonal) structure.[1–3] Rutile GeO₂, in particular, is chemically and structurally similar to other oxide materials such as SnO₂ and TiO₂, two common materials in the transparent conducing oxide (TCO) and semiconductor industries which also crystallize in the rutile phase, among multiple polytypes.[4–6] Rutile belongs to the tetragonal Bravais lattice group with lattice parameters $a = b \neq c$ and $\alpha = \beta = \gamma = 90°$. The experimental lattice parameters of rutile GeO₂ are $a = 4.4066$ Å and $c = 2.8619$ Å.[1] The crystal structure is shown in Fig. 1. Considering the anisotropy of the crystal structure, measurements are often performed along two axes, $\perp \vec{c}$ (i.e. along $\vec{a}$) and $\parallel \vec{c}$. One main issue that rutile GeO₂ could overcome is the dearth of $p$-type TCO materials, as many oxides exhibit flat valence bands that give rise to trapped hole polarons.[7,8]

---

a) Author to whom correspondence should be addressed. Electronic mail: kioup@umich.edu.

While $TiO_2$ and $SnO_2$ are difficult to *p*-type dope, rutile $GeO_2$ has recently been theoretically predicted to be ambipolarly dopable. Chae *et al.* showed through first-principles defect calculations that group-III metals such as Al on the Ge site are possible acceptors with rather large ionization energies of 0.45-0.54 eV, yet the codoping of group-III dopants with H and subsequent annealing allows the incorporation of high acceptor concentrations that enable *p*-type conduction through the impurity band.[9] The largest experimental band gap of rutile $GeO_2$ (4.68 eV)[10] is also wider than that of $TiO_2$ and $SnO_2$ (3.03 eV and 3.6 eV, respectively),[8,11–15] and falls in the UVC region of the electromagnetic spectrum (4.28-6.20 eV; 200-290 nm). This gap value is desirable for solar-blind photodetectors, which require materials that absorb wavelengths shorter than 290 nm while maintaining transparency at lower photon energies. Additionally, the UVC region is the most effective for germicidal applications, e.g. water purification and food sterilization.[16] The two properties of an ultra-wide-band gap in the UVC and the possibility of ambipolar doping motivate the study of rutile $GeO_2$ for optoelectronic applications.

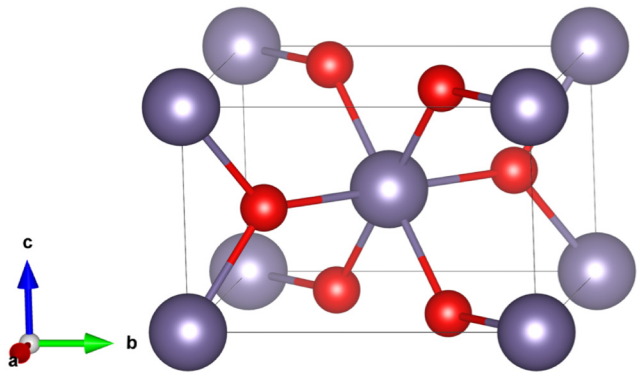

FIG. 1. Crystal structure of r-$GeO_2$ (large grey atoms are Ge and small red atoms are O). Our calculations employed experimental lattice parameters and atomic positions from Ref. 1.

The majority of previous work on GeO$_2$ focuses on the $\alpha$ (quartz) polytype. Also, most theoretical work on rutile GeO$_2$ does not go beyond density functional theory (DFT)[2,17–19] and suffers from the band-gap underestimation problem. Recently, Chae *et al*. and Samanta *et al*. reported band structures calculated with hybrid functionals and many-body perturbation theory (GW method), respectively.[9,20] Samanta *et al* showed that varying the lattice parameters of r-GeO$_2$ significantly impacts the calculated value of the band gap, even more so than the type of functional used on the DFT level.[20] Even on the G$_0$W$_0$ level, the band gap can range from 4.05 eV to 5.78 eV by decreasing the volume from 55.91 Å$^3$ to 51.03 Å$^3$, all of which maintain a dynamically stable structure. Aside from the crystal and electronic structures, several studies have measured[3,10] or calculated[17,21] the complex dielectric function, refractive index, absorption coefficients, and other related properties. The theoretical reports in the literature remain on the DFT level, which are affected by the gap underestimation and are not accurate for comparison to experiment.

In this work, we apply electronic-structure calculations based on many-body perturbation theory to characterize the electronic and optical properties of rutile GeO$_2$ (r-GeO$_2$) and enable accurate comparisons with experiment. We determine the band structure, band gap, and carrier effective masses. We explore the formation of self-trapped hole polarons to determine their possible impact on hole transport. We also calculate the interband optical matrix elements along both crystallographic directions and analyze the imaginary part of the dielectric function. We also obtain the real part of the dielectric function, refractive index, and absorption coefficient along both crystallographic directions. Our work provides atomistic insights on experimental measurements and can guide future studies on deep-UV optoelectronic and solar-blind photodetection applications.

## II. COMPUTATIONAL METHODS

Our first-principles calculations are based on density functional and many-body perturbation theories. The experimental lattice parameters of r-GeO$_2$ ($a$ = 4.4066 Å and $c$ = 2.8619 Å)[1] were used for the electronic and optical properties calculations without structural relaxation to allow a direct comparison to experiment. First, the mean-field charge density, wave functions, and band eigenvalues were calculated within the local-density approximation (LDA)[22,23] with the Quantum ESPRESSO code[24] before applying the G$_0$W$_0$ method with the BerkeleyGW software[25] to include electron self-energy effects on the band structure. Norm-conserving pseudopotentials[26] were used to describe the interactions of the valence electrons of the Ge ($4s$ and $4p$) and O ($2s$ and $2p$) atoms with the ionic cores. We employed a plane-wave basis set with a 90 Ry cutoff energy for the DFT wave-function calculations and sampled the Brillouin zone (BZ) with an 8×8×12 grid, which converge the total energy of the system to within 1 mRy/atom. For the G$_0$W$_0$ calculations we used a plane-wave basis for the dielectric matrix up to a 35 Ry cutoff energy and sampled the BZ with a 4×4×6 grid. We summed over 2,000 and 2,504 total bands for the dielectric-matrix and the self-energy calculations, respectively. The generalized plasmon-pole model[27] and the static-remainder approach[28] were incorporated to extrapolate the dielectric function to finite frequency and more rapidly converge the Coulomb-hole summation over unoccupied states, respectively. The convergence errors stemming from our choice of G$_0$W$_0$ calculation parameters are ~5 meV from the total bands of the dielectric-matrix calculation, ~10 meV from the total bands of the self-energy calculation,

~50 meV from the screened Coulomb cutoff of 35 Ry, and ~10 meV from the BZ sampling grid of 4×4×6, resulting in a total estimated convergence error for our $G_0W_0$ calculations of ~75 meV.

The DFT and $G_0W_0$ band structures, as well as the velocity-operator matrix elements (VMEs)

that describe band-to-band optical transitions,[29] were interpolated to fine BZ sampling grids (up

to 160×160×240) with the maximally localized Wannier function (MLWF) method[30] and the

wannier90 code[31] to determine the imaginary part of the dielectric function, $\varepsilon_2$.[32] The Bethe-

Salpeter-equation (BSE) method as implemented in BerkeleyGW was also used to calculate $\varepsilon_2$

on a 16×16×24 BZ sampling grid, including twelve valence and two conduction bands, and using

0.2 eV Gaussian broadening to evaluate excitonic effects on the dielectric function. Self-trapped

hole polarons are investigated using the Heyd-Scuseria-Ernzerhof (HSE06) functional, with the

Hartree-Fock mixing parameter, BZ sampling, and the supercells described in Ref. 9. For the polaron calculations only we relaxed the crystal structure. Our relaxed lattice parameters ($a = 4.394$ Å and $c = 2.866$ Å) agree with experiment[1] within <0.3%. We evaluate the stability of self-trapped holes by evaluating the self-trapping energy ($E_{ST}$) as defined in Ref. 8.

## III. ELECTRONIC PROPERTIES

### A. Band structure and band gap

The DFT-LDA and $G_0W_0$ band structures of r-GeO$_2$ (Fig. 2) provide information about

its fundamental electronic and optical properties. Both the DFT-LDA and $G_0W_0$ band structures

exhibit direct fundamental band gaps at Γ. However, as shown in Table I, the DFT band gap is

only 1.96 eV, which is 56% smaller than the $G_0W_0$ band gap of 4.44 eV and 58% smaller than

the experimental value of 4.68 eV.[10] Such a large difference is expected from the DFT

calculation and exemplifies the need to include many-body effects; the $G_0W_0$ band gap differs by only 5% from experiment. Previous calculations by Chae *et al.* using HSE06 with 35% Hartree-Fock exchange found a band gap of 4.64 eV after relaxing the structure and obtaining lattice parameters $a$ = 4.394 Å and $c$ = 2.866 Å.[9] Samanta *et al.* investigated the pressure dependence of the band gap of r-GeO$_2$ and found that the structure is dynamically stable over a wide range of lattice volumes from -14% to +3% strain, referenced to the experimental volume of 55.57 Å$^3$.[1,20] They calculate and report several G$_0$W$_0$ band gaps from 4.05 eV to 5.78 eV using different unit-cell volumes reported in the literature, ranging from 51.03 Å$^3$ to 55.91 Å$^3$.[1,2,20] For this reason we performed our calculations using the experimental lattice parameters to avoid artifacts from the underestimation of the lattice constants by DFT-LDA.

**B. Effective masses and polaron properties**

We also obtain the hole and electron effective masses by fitting the hyperbolic equation:

$$E(k) = \frac{\mp 1 \pm \sqrt{1 + 4\alpha\hbar^2 k^2/(2m^*)}}{2\alpha} + E_0, \qquad (1)$$

to the Wannier-interpolated G$_0$W$_0$ band structure along the $\Gamma \rightarrow X$ and $\Gamma \rightarrow Z$ directions for the top valence band (VB) and bottom conduction band (CB). The energy ranges from the corresponding band extrema used to perform the effective-mass fits and the associated wave-vector ranges are 0.5 eV for $m^*_{e\perp}$ (17.5% of the $\Gamma \rightarrow X$ BZ segment), 0.8 eV for $m^*_{e\parallel}$ (11.4% of $\Gamma \rightarrow Z$), 0.4 eV for $m^*_{h\perp}$ (28.1% of $\Gamma \rightarrow X$ BZ), and 0.3 eV for $m^*_{h\parallel}$ (18.2% of $\Gamma \rightarrow Z$ BZ). In Eq. 1, $E(k)$ is band energy as a function of crystal momentum $k$, $\alpha$ is the non-parabolicity fitting parameter, $\hbar$ is the reduced Planck constant, $m^*$ is the electron (-/+) or hole (+/-) effective mass, and $E_0$ is the energy of the VBM (-/+) or CBM (+/-). While the band gap changes significantly between DFT and G$_0$W$_0$, the curvatures of the bands remain approximately the same between the two levels of theory, especially at Γ. As such, the hole and electron effective masses calculated from both methods agree well, so we only report the $G_0W_0$ effective masses in Table II. The electron effective mass along Γ → X ($m_{e\perp}^* = 0.43\ m_0$) is about a factor of 2 larger than along Γ → Z ($m_{e\parallel}^* = 0.23\ m_0$), but both are comparable to those of other wide-band-gap materials; rutile $SnO_2$ (r-$SnO_2$), a chemical and structural analogue to r-$GeO_2$, has essentially the same $m_{e\parallel}^*$ (0.234 $m_0$) and slightly lighter $m_{e\perp}^*$ (0.299 $m_0$).[33] Yan *et al.* reported GaN electron effective masses of $m_{e\parallel}^* = 0.19\ m_0$ and $m_{e\perp}^* = 0.21\ m_0$.[34] The electron effective mass of β-$Ga_2O_3$ ranges from ~0.23 $m_0$ to 0.34 $m_0$ depending on the crystallographic direction.[35,36] On the other hand, the hole effective mass of r-$GeO_2$ is notably small for such an ultra-wide-gap material. Our calculations show values of $m_{h\perp}^* = 1.28\ m_0$ and $m_{h\parallel}^* = 1.74\ m_0$. For comparison, for r-$SnO_2$ Schleife *et al.* report $m_{h\perp}^* = 1.21\ m_0$ and $m_{h\parallel}^* = 1.47\ m_0$,[7] while Varley *et al.* report $m_{h\perp}^* = 1.37\ m_0$ and $m_{h\parallel}^* = 1.61\ m_0$.[37] Other common ultra-wide-gap materials like AlN and β-$Ga_2O_3$ have significantly larger hole effective masses. For AlN, the heavy hole effective masses are $m_{hh\perp}^* = 10.42\ m_0$ and $m_{hh\parallel}^* = 3.53\ m_0$.[38] The valence bands of β-$Ga_2O_3$ are notoriously flat[39], resulting in a large hole effective mass of ~40 $m_0$ [40–42] that yields self-trapped hole polarons[8]. However, for r-$GeO_2$, we calculate the hole self-trapping energy ($E_{ST}$) to be 0.0094 eV, indicating that in the absence of impurities self-trapped hole polarons are only weakly bound and thermally dissociate to delocalized holes at room temperature (hole polarons in r-$GeO_2$ are more strongly bound to negatively-charged acceptors such as Al, however[9]). The delocalized nature of holes stems from their light effective mass and separates r-$GeO_2$ from other wide-gap oxides and is one reason why this material can also be *p*-type doped. Overall, the band structure provides the fundamental information about the electronic properties of this material such as its direct band gap of 4.44 eV and the small, anisotropic electron and hole effective masses, while it also serves as the starting point for investigating optical properties.

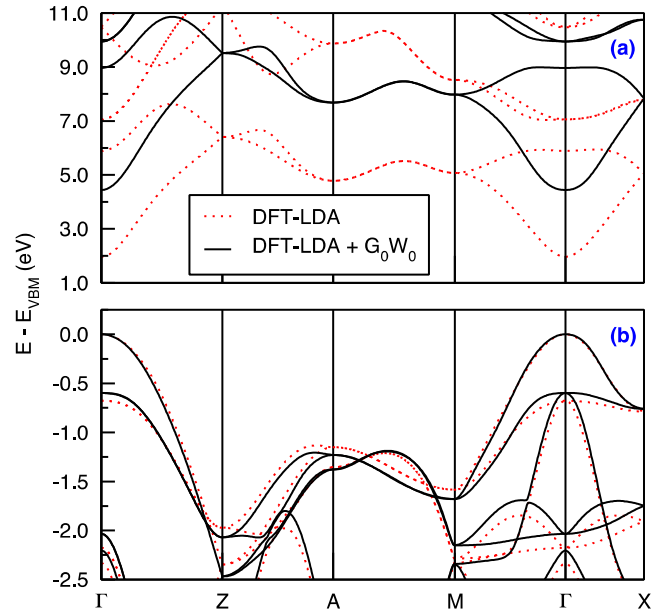

FIG. 2. (a) Bottom conduction and (b) and top valence bands of r-$GeO_2$ within DFT-LDA (red dotted) and DFT-LDA + $G_0W_0$ (black solid). The band gap is direct at $\Gamma$ with a magnitude of 1.96 eV (LDA) and 4.44 eV ($G_0W_0$).

TABLE I. Magnitude of the direct band gap of r-$GeO_2$ at $\Gamma$ calculated with different methods and compared to experimental optical measurements. The four calculations shown used the same set of experimental lattice parameters: $a = 4.4066$ Å and $c = 2.8619$ Å.[1]

| Method | Band Gap (eV) |
|---|---|
| DFT-PBE[20] | 1.80 |
| DFT-LDA (present work) | 1.96 |
| DFT-PBE + $G_0W_0$[20] | 4.20 |
| DFT-LDA + $G_0W_0$ (present work) | 4.44 |
| Experiment (UV-absorption)[10] | 4.68 |

TABLE II. Effective masses of electrons and holes of r-GeO$_2$ along the $\Gamma \rightarrow X$ and $\Gamma \rightarrow Z$ directions calculated by a hyperbolic fit to the G$_0$W$_0$ band structure.

| Direction | $m_e^*$ $(m_0)$ | $m_h^*$ $(m_0)$ |
|-----------|-----------------|-----------------|
| $\Gamma \rightarrow X$ | 0.43 | 1.28[a] |
| $\Gamma \rightarrow Z$ | 0.23 | 1.74[a] |

[a] values previously reported in Ref. 9

## IV. OPTICAL PROPERTIES

### A. Absorption onsets

We determined the character (allowed or forbidden) of optical transitions across the band gap of r-GeO$_2$, that ultimately control its optical spectra, by analyzing the band energies and interband VMEs. We examine the VMEs for transitions from the top six valence bands to the bottom conduction band at $\Gamma$ (Table III), as these are the only bands that contribute to optical absorption for photon energies up to 10 eV, and find several noteworthy features. First, the VME is zero along both the $\perp \vec{c}$ and $\parallel \vec{c}$ directions for transitions from the top valence band to the bottom conduction band. This coincides with the small value for the imaginary part of the dielectric function $\varepsilon_2$ (plotted in logarithmic scale in Fig. 3) at photon energies near the band gap energy. We also show the VMEs for the top valence band to bottom conduction band transition for electron wave vectors along the $\Gamma \rightarrow X$ and $\Gamma \rightarrow Z$ BZ directions in Fig. 4. While the $\vec{x}$ Cartesian component ($\perp \vec{c}$) of the VMEs near $\Gamma$ is zero along $\Gamma \rightarrow Z$ and increases rapidly along $\Gamma \rightarrow X$ , the $\vec{z}$ component ($\parallel \vec{c}$) VME remains negligible along both directions. This explains the stronger absorption onset for light polarized along the $\perp \vec{c}$ direction (Fig. 3), which is determined by the interband VMEs in the vicinity of $\Gamma$. Ultra-violet absorption measurements performed by Stapelbroek *et al.* also pointed to a forbidden transition for the fundamental band gap of r-GeO$_2$,[10] and our calculations are in agreement with this previous analysis. For the second and third valence bands, which are degenerate at $\Gamma$, the VME is strong $\perp \vec{c}$ but not $\parallel \vec{c}$ (Table III). Therefore, our calculations reveal that r-GeO$_2$ only absorbs weakly for photon energies between the band gap (4.44 eV) and the energy difference between the second valence band and the CBM (5.04 eV) due to the fundamental interband transition being dipole forbidden. The value for $\varepsilon_2 \perp \vec{c}$ plateaus with increasing photon energy above 5.04 eV (Fig. 3). Transitions for light polarized along $\vec{c}$ from the top five valence bands to the CBM are weak, with $\varepsilon_2$ increasing slowly until the sixth valence band to CBM transition energy is reached at 6.65 eV, above which r-GeO$_2$ absorbs strongly also along $\vec{c}$.

TABLE III. Energies and matrix elements of optical transitions from the top six valence bands to the bottom conduction band of r-GeO$_2$ at $\Gamma$, calculated from the G$_0$W$_0$ band structure. The magnitudes of the velocity-operator matrix elements (in Hartree atomic units) for each band-to-band transition are shown for both the $\perp \vec{c}$ and $\parallel \vec{c}$ directions.

| Valence Band Index | $E_{CB} - E_{VB,i}$ (eV) | VME ($\perp \vec{c}$) | VME ($\parallel \vec{c}$) |
|---|---|---|---|
| 1 | 4.44 | 0.00 | 0.00 |
| 2 | 5.04 | 0.53 | 0.00 |
| 3 | 5.04 | 0.53 | 0.00 |
| 4 | 6.47 | 0.00 | 0.00 |
| 5 | 6.47 | 0.00 | 0.00 |
| 6 | 6.65 | 0.00 | 0.96 |

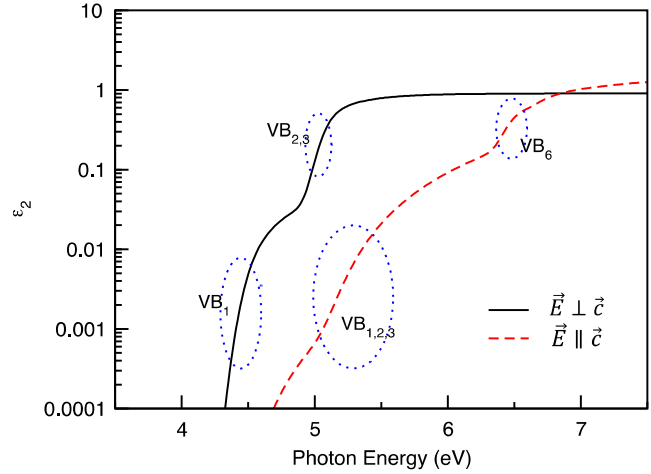

FIG. 3. Imaginary part of the dielectric function of r-GeO$_2$ in logarithmic scale calculated using the maximally localized Wannier function method on a fine 160×160×240 BZ sampling grid. The spectra show multiple electronic transition onsets with increasing photon energy. The approximate onsets of specific valence-to-conduction band transitions are highlighted. While the fundamental band gap occurs at 4.44 eV, the corresponding optical transition is dipole-forbidden, resulting in a small value for $\varepsilon_2$ at that photon energy.

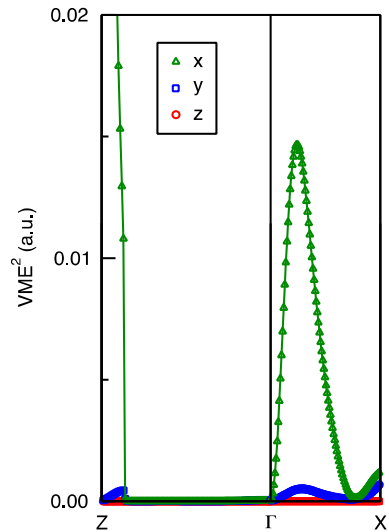

FIG. 4. Velocity matrix elements squared (in Hartree atomic units) for electron wave vectors along the $\perp \vec{c}$ (Γ→X) and $\parallel \vec{c}$ (Γ→Z) directions of the BZ. Only the Γ→X direction shows strong

VMEs for light polarized $\perp \vec{c}$ near $\Gamma$, which explains the stronger absorption onset for the $\perp \vec{c}$ polarization in Fig. 3.

### B. Excitonic effects

We also calculated excitonic effects on the imaginary part of the dielectric function using the BSE method, and the results are shown in Fig. 5. Due to the computational cost of the BSE calculations, we used a coarser 16×16×24 BZ sampling grid to show the overall difference of the $\varepsilon_2$ spectra both with and without excitonic effects included (in comparison, the spectra calculated without excitonic effects in Fig. 3 were obtained for a finer BZ sampling grid and exhibit sharper spectral features). The $\varepsilon_2$ spectra are converged for this BZ sampling grid. The general shape and onsets of the $\varepsilon_2$ spectra calculated both with and without excitonic effects are similar. As with the r-SnO$_2$ $\varepsilon_2$ results,[7] we find that r-GeO$_2$ has anisotropic optical properties for light polarized along different crystallographic orientations, with the onset of absorption for $\vec{E} \perp \vec{c}$ at lower energies than $\vec{E} \parallel \vec{c}$ as discussed previously. Excitonic effects introduce an overall spectral-weight shift to lower energies. The $\varepsilon_2$ curves including excitonic effects rise much quicker than those without excitonic effects. These excitonic corrections to the optical spectra do not originate from a shifting of the peak positions by the exciton binding energy (which is only of the order of 39 meV as we discuss below), but rather from the modification of the optical matrix elements due to the coherent coupling of excited states induced by electron-hole interactions.[43]

We also investigate the exciton binding energies for the lowest exciton states. Convergence for the fine k-point sampling grid within BSE are shown in Table SI and Fig. S1 in the Supplementary Material. Our calculations using a 16×16×24 BZ sampling grid result in an exciton binding energy of 184 meV for $\vec{E} \perp \vec{c}$ and 169 meV for $\vec{E} \parallel \vec{c}$, which significantly overestimate the experimental value for $\vec{E} \perp \vec{c}$ (39 meV).[10] We extrapolate a value of 152 meV for $\vec{E} \perp \vec{c}$ and an infinite BZ sampling grid (Fig. S2). A similar result and discussion were presented by Schleife et al. for r-SnO$_2$, where they discuss the importance of ionic screening that is typically omitted in GW and BSE first-principles calculations.[7] Schleife et al. report exciton binding energies for r-SnO$_2$ of 222 meV for $\vec{E} \perp \vec{c}$ and 191 meV for $\vec{E} \parallel \vec{c}$, closely matching the values we calculate for r-GeO$_2$. While GW and BSE methods use the high-frequency dielectric constant $\varepsilon_\infty$ which does not include ionic screening effects, the static dielectric constant does include ionic screening. By using the average static dielectric constants in the formula for the Wannier-Mott exciton, they estimate exciton binding energies equal to 19 meV ($\vec{E} \perp \vec{c}$) and 16 meV ($\vec{E} \parallel \vec{c}$), which are in much better agreement with the experimental value for SnO$_2$ of ~30 meV[44,45]. Using the same Bohr model where $E_b = 13.6$ eV $\times \mu^*/\varepsilon_0^2$ (where $\mu^*$ is the reduced effective mass, and $\varepsilon_0$ is the static dielectric constant) and our previously calculated static dielectric constants $\varepsilon_{0,\perp\vec{c}} = 16.02$ and $\varepsilon_{0,\parallel\vec{c}} = 7.78$,[9] we estimate exciton binding energies of 17 meV ($\vec{E} \perp \vec{c}$) and 46 meV ($\vec{E} \parallel \vec{c}$). The reduced effective masses were calculated in both directions using the values from Table II, where $\mu^*_{\perp\vec{c}} = 0.32$ and $\mu^*_{\parallel\vec{c}} = 0.20$. If we instead use the directionally averaged static dielectric constant 13.27 rather than the direction-dependent values, we obtain $E_b = 25$ meV ($\vec{E} \perp \vec{c}$) and 16 meV ($\vec{E} \parallel \vec{c}$). Therefore, the inclusion of ionic screening brings our calculated values in better agreement with experiment (39 meV).[10]

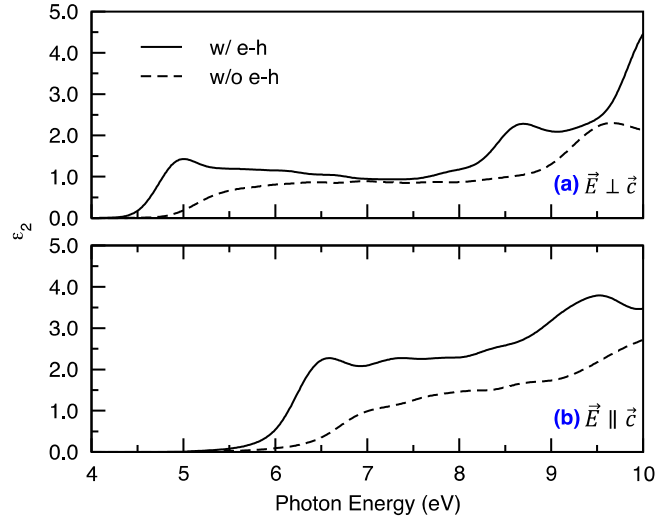

FIG. 5. Imaginary part of the dielectric function of r-GeO$_2$ as calculated with the BSE on a 16×16×24 BZ sampling grid without (dashed) and with (solid) exitonic effects included. Panel (a) shows the results for light polarized perpendicular to $\vec{c}$ and (b) parallel to $\vec{c}$.

**C. Optical constants**

We further utilized the Kramers-Kronig relation to obtain the real part of the dielectric function ($\varepsilon_1$) of r-GeO$_2$ from the imaginary part ($\varepsilon_2$) and hence determine the refractive index ($n$), extinction coefficient ($\kappa$), and absorption coefficient ($\alpha$) over the 0-10 eV photon-energy range. To ensure convergence of the real part in this energy range, the imaginary part was calculated and the Kramers-Kronig relation integrated for energies up to 45 eV. For this calculation, we used the $\varepsilon_2$ without excitonic effects calculated from maximally localized Wannier functions on a 160×160×240 BZ sampling grid. While we did explore excitonic corrections to the imaginary dielectric function for photon energies up to 10 eV, the Kramers-Kronig transformation requires a much wider energy range to yield accurate values for the real part of the dielectric function (and, subsequently, the refractive index and the absorption coefficient), which is computational demanding. Figure 6 shows both components of the dielectric function from 0 eV to 10 eV. The refractive index $n$ and extinction coefficient $\kappa$ are determined according to:

$$n = \sqrt{\left(\sqrt{\varepsilon_1^2 + \varepsilon_2^2} + \varepsilon_1\right)/2} \qquad (2)$$

and:

$$\kappa = \sqrt{\left(\sqrt{\varepsilon_1^2 + \varepsilon_2^2} - \varepsilon_1\right)/2} \; . \qquad (3)$$

The calculated refractive index is quite large, ranging from ~1.9 to ~2.5 between 0 eV to 10 eV. Our results agree well with those calculated using ultrasoft pseudopotentials within the generalized gradient approximation on the DFT level by Liu *et al.*[17] The refractive indices of other well-known visible transparent semiconductors are 1.46-1.51 from 2.3 eV to 5.2 eV (SiO$_2$ glass)[13] and 2.36-2.78 from 1.2 eV to 3.4 eV (w-GaN)[46], for example. Between 1.77-3.10 eV (700 nm to 400 nm), the refractive index of r-GeO$_2$ changes by only 0.03 for both crystallographic directions. Such a small change in $n$ over the entire visible range is one sign that r-GeO$_2$ is a promising candidate for applications requiring low chromatic dispersion. For a direct comparison to experiment, we determined the absorption coefficient $\alpha$ via:

$$\alpha(E) = \frac{4\pi\kappa E}{hc}, \qquad (4)$$

where $E$ is the photon energy, $h$ is the Planck constant, and $c$ is the speed of light. Figure 7 shows the calculated refractive index and absorption coefficient as a function of energy. Our calculations confirm the previous experimental finding by Stapelbroek *et al.* that the first band-to-band transition is dipole-forbidden and agree with the dichroism measured in the polarized edge absorption experiment.[10] This anisotropy is apparent in the absorption coefficient shown in Fig. 7(b); $\alpha$ becomes sizable only above ~5 eV, the energy of the second VB to CBM transition. Although our calculated $G_0W_0$ band gap is 0.24 eV smaller than that reported experimentally, when we rigidly shift our absorption coefficient curves by this amount to align the band gaps, the calculated results closely match the 77 K experimental data, as shown in Fig. 8.[10] We find that, just like the experiment, $\alpha$ for $\vec{E} \perp \vec{c}$ has a much steeper slope than for $\vec{E} \parallel \vec{c}$ at energies near the band gap. From the lack of strong VMEs of electronic transitions within 2 eV of the band gap along $\vec{c}$, this behavior is expected. Overall, the electronic and optical properties derived from the $G_0W_0$ band structure agree well with experimental results.

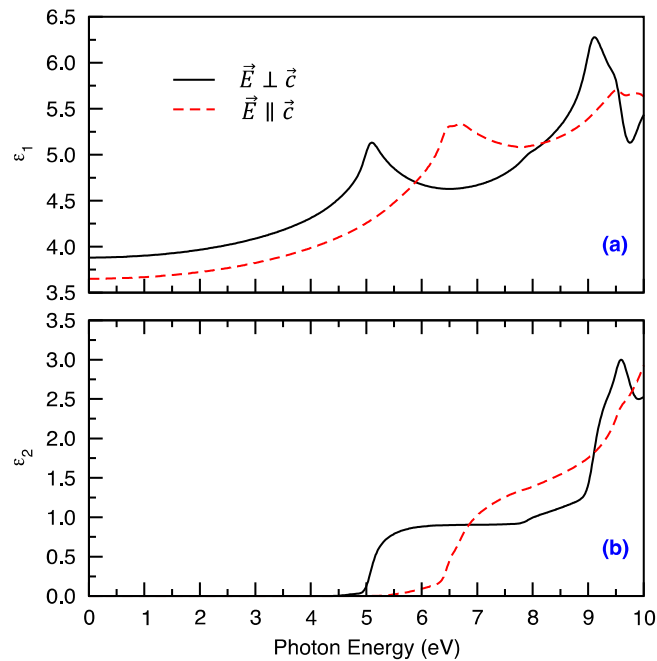

FIG. 6. (a) Real and (b) imaginary part of the dielectric function of r-GeO$_2$ for electric-field polarizations along the two main crystallographic directions.

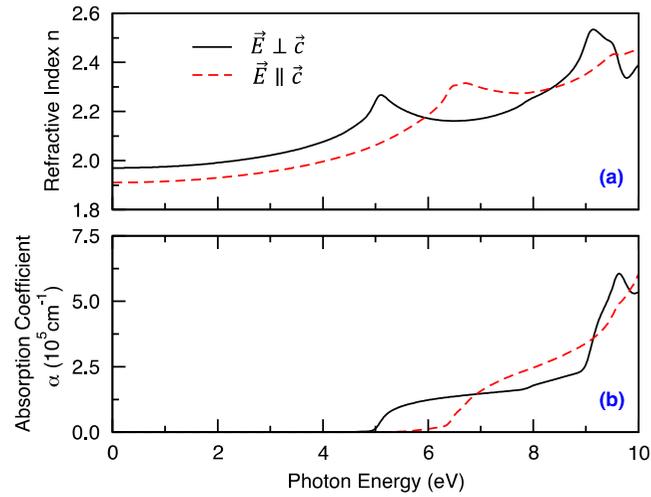

FIG. 7. (a) Refractive index and (b) and absorption coefficient for the two main crystallographic directions of r-GeO$_2$.

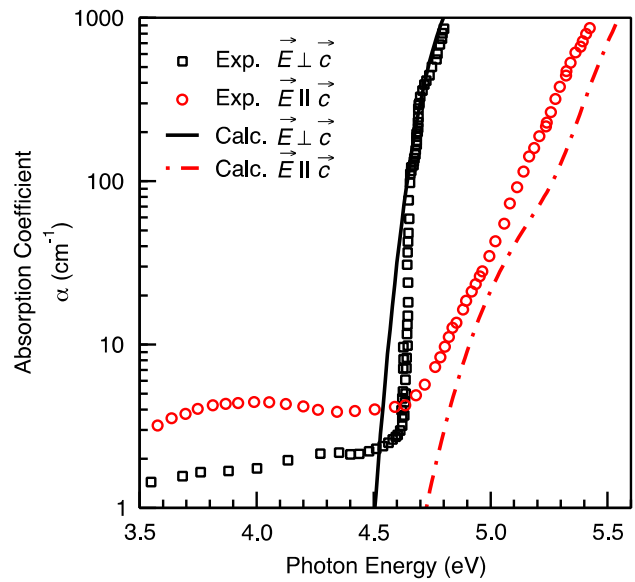

FIG. 8. Comparison of our calculated (0 K simulation) absorption coefficient for r-GeO$_2$ along $\vec{E} \perp \vec{c}$ and $\vec{E} \parallel \vec{c}$ to experimental data from Stapelbroek *et al*. collected at 77 K (Ref. 10).

### D. Applications in solar-blind photodetection

The optical absorption of r-GeO$_2$ is important to consider for applications in solar-blind photodetection. An ideal solar-blind photodetector absorbs wavelengths shorter than 290 nm (4.28 eV) but is entirely transparent in the visible and near-UV ranges. One promising solar-blind material is β-Ga$_2$O$_3$, which has a wide band gap (~4.5 eV) close to that of r-GeO$_2$.[39,47–49] β-Ga$_2$O$_3$ nanodevices have been shown to selectively absorb light in the desired solar-blind range. An advantage of nanoscale devices is the potential miniaturization of photodetectors, which opens future work for r-GeO$_2$ devices. L. Li *et al*. fabricated bridged β-Ga$_2$O$_3$ nanobelts that have a high response at 250 nm that is ~10$^6$ higher than in the visible,[48] indicating an excellent solar-blind photoresponse for this material. To compare β-Ga$_2$O$_3$ directly to r-GeO$_2$, we examine their absorption coefficients at energies that are important for solar-blind applications. In the case of an annealed β-Ga$_2$O$_3$ film, the absorption coefficient ranges from ~1-4×10$^4$ cm$^{-1}$ for energies below the band gap and increases to ~2×10$^5$ cm$^{-1}$ at 5 eV (248 nm), for instance.[50] The cause of sub-band-gap absorption is the presence of defects and impurities.[50,51] In the present calculations on r-GeO$_2$ for the pristine, bulk crystal, $\alpha = 0$ cm$^{-1}$ for energies below the band gap. At 5 eV (0.56 eV above the calculated band gap), $\alpha = $ ~3×10$^3$ cm$^{-1}$ for $\vec{E} \perp \vec{c}$, which is quite weak compared to the absorption coefficient at just 0.5 eV higher (~8×10$^4$ cm$^{-1}$). The concentration of defects, impurities, and free carriers should be studied experimentally to better understand the ratio $\alpha_{E>4.28\,eV}/\alpha_{E<4.28\,eV}$ in real r-GeO$_2$ devices with defects and impurities for the purpose of solar-blind applications. An interesting similarity between β-Ga$_2$O$_3$ and r-GeO$_2$ is that each has an intrinsic feature in the band structure that suppresses optical absorption at the fundamental band gap. In the case of β-Ga$_2$O$_3$, the fundamental band gap is indirect, so absorption is weak until ~30 meV above the band gap when the energy of the direct gap is reached.[39] As we have already shown for r-GeO$_2$, the electronic transition from the VBM to the CBM is forbidden, so absorption is suppressed for a 600 meV energy range until the second VB to CBM energy is reached. This behavior is notably different for dipole-allowed direct-gap materials such as GaN, for which the absorption coefficient increases strongly for energies above the band gap.[52] In contrast, the weak interband matrix elements of r-GeO$_2$ shift the absorption onset further into the UVC range, away from the solar-blind threshold of 290 nm. While β-Ga$_2$O$_3$ absorbs more strongly than r-GeO$_2$ since the first several optical transitions are dipole allowed, r-GeO$_2$ is more promising for solar-blind photodiodes because it can be ambipolarly doped and has lighter hole effective masses, which is indicative of high hole mobility.

## V. CONCLUSIONS

In summary, we applied first-principles calculations to analyze the band structure, interband transitions, and optical properties in r-GeO$_2$ for potential applications in optoelectronic devices. We find a direct band gap of 4.44 eV at $\Gamma$, while the carrier effective masses are similar to or smaller than other wide-band-gap semiconductors. Unlike many other wide-band-gap oxides, delocalized holes are found to be energetically more stable than self-trapped hole polarons owing to their small effective mass. Our estimated exciton binding energies are in good agreement with experiment. The optical matrix elements reveal that the transition from the top valence band to the conduction band minimum at $\Gamma$ is forbidden, while the transition from the second VB to CBM is allowed for $\vec{E} \perp \vec{c}$ only. Absorption occurs for $\vec{E} \parallel \vec{c}$ starting at a higher energy of 6.65 eV. The short absorption onset wavelength of r-GeO$_2$ in addition to its semiconducting properties and ambipolar dopability support its applications as a UV-transparent conductor and in solar-blind photodetector devices.

## SUPPLEMENTARY MATERIAL

Additional details related to the convergence of excitonic properties (imaginary part of the dielectric function and exciton binding energy with respect to the fine BSE grid) can be found in the Supplementary Material.

## ACKNOWLEDGEMENTS

This work was supported by the Designing Materials to Revolutionize and Engineer our Future (DMREF) Program under Award No. 1534221, funded by the National Science Foundation. K.A.M. also acknowledges the National Science Foundation for support from Graduate Research Fellowship Program through Grant No. DGE 1256260. This research used resources from the National Energy Research Scientific Computing Center, a DOE Office of Science User Facility supported by the Office of Science of the U.S. Department of Energy through Grant DE-AC02-05CH11231.

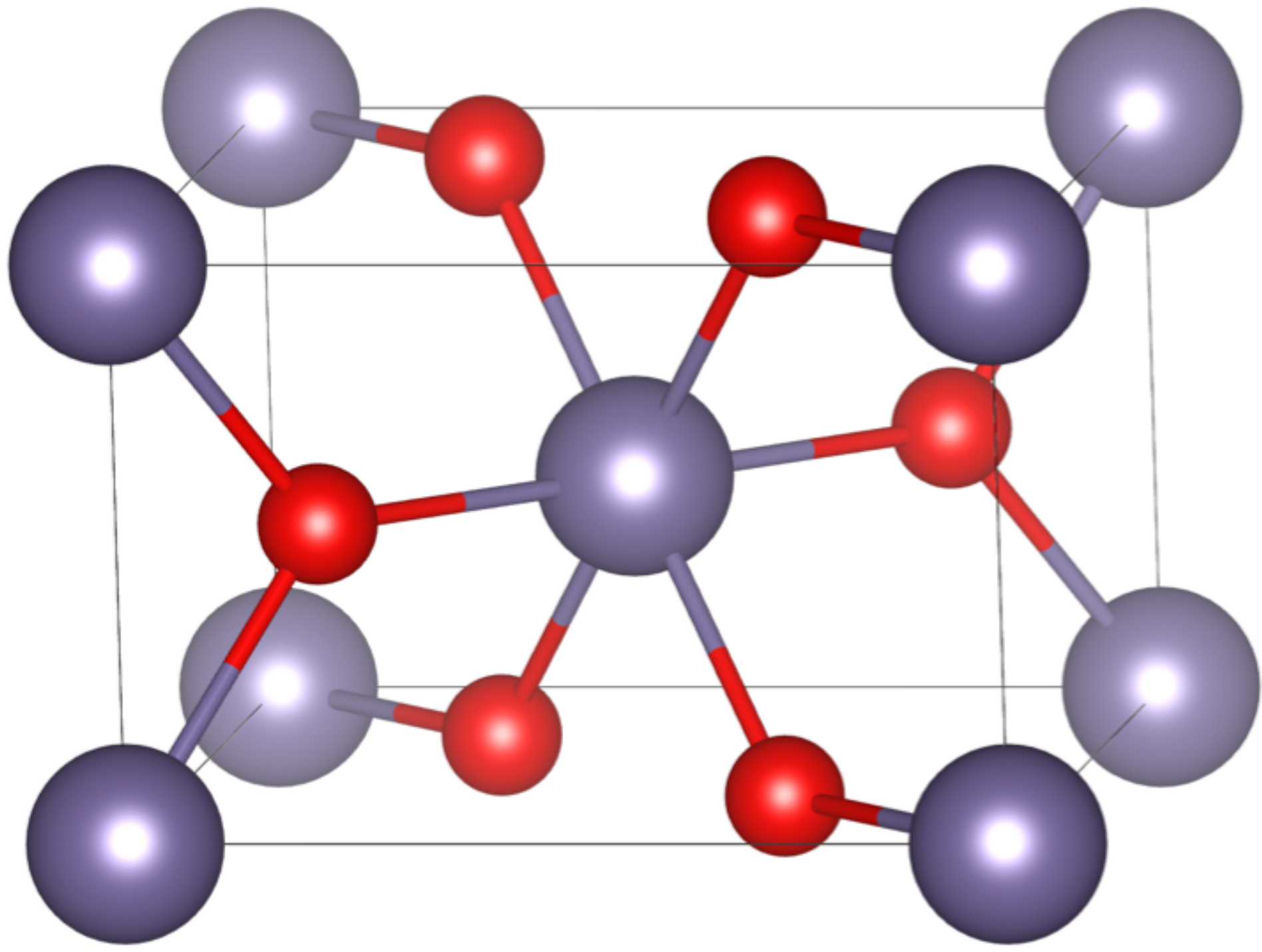

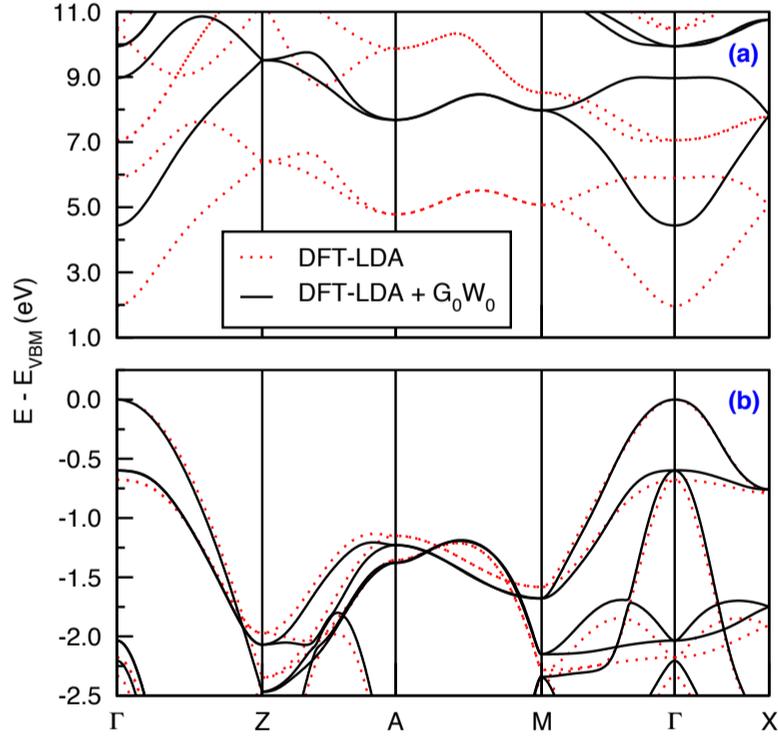

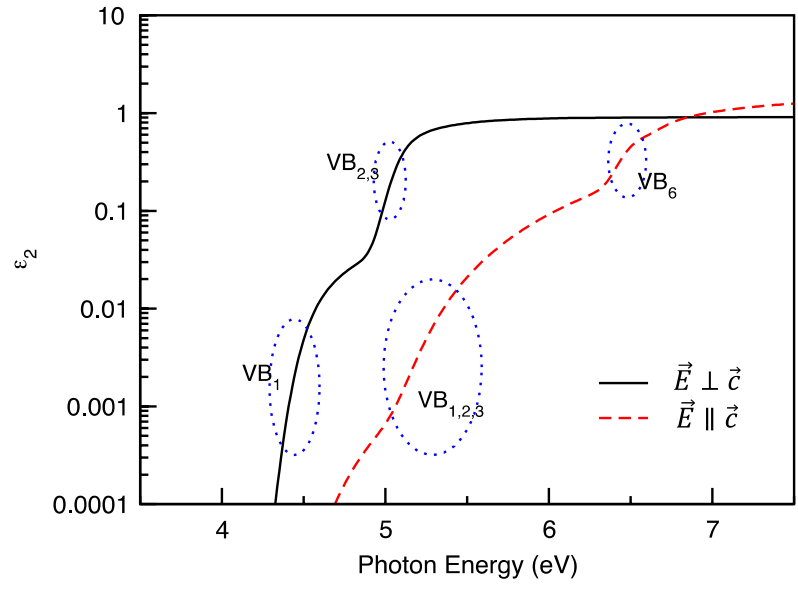

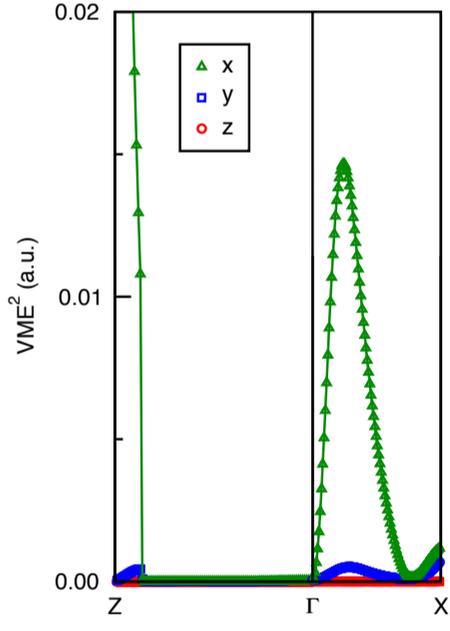

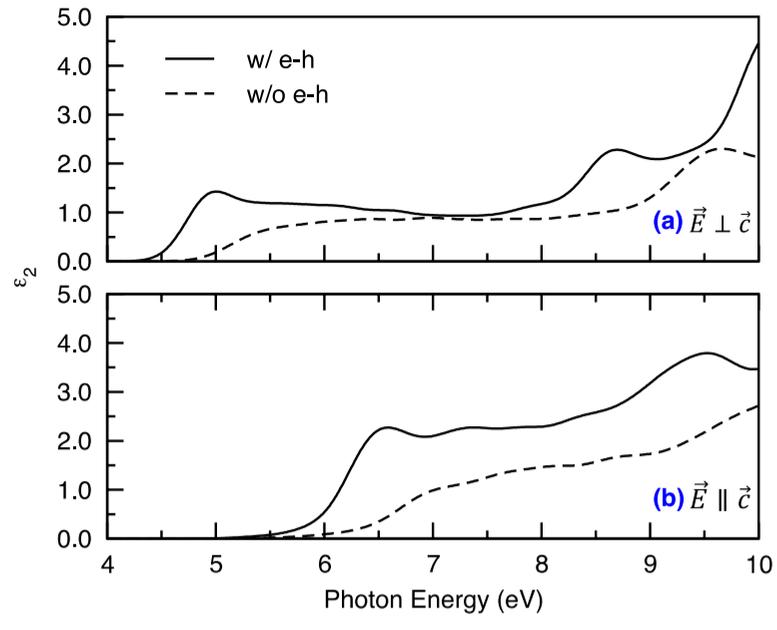

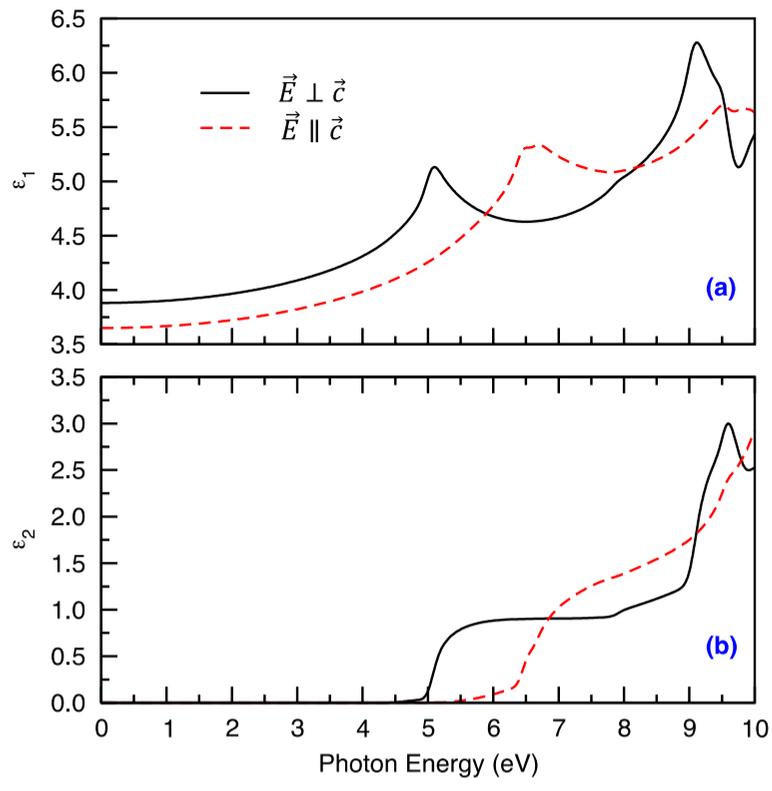

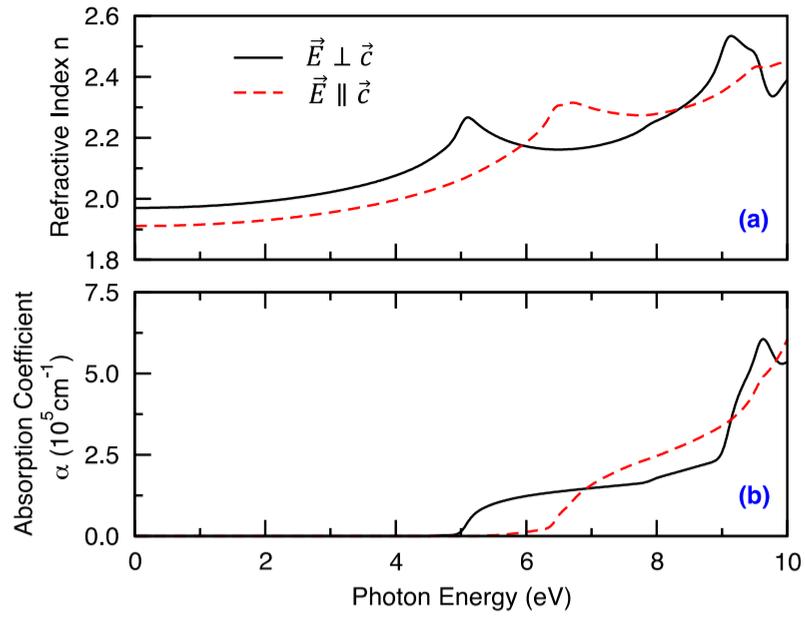

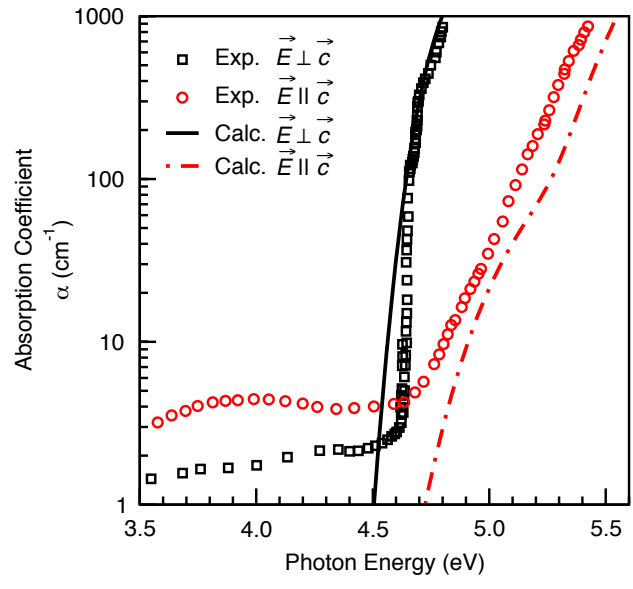